\documentclass[aps,pra,twocolumn,showpacs,floatfix]{revtex4}
\usepackage[usenames]{color}
\usepackage{amsmath}
\usepackage{graphicx}
\usepackage{amsfonts}
\usepackage{euscript}
\usepackage{subfigure}
\usepackage{color}
\begin{document}
\bibliographystyle{apsrev}
\title{Transient Impulsive Electronic Raman Redistribution}
\author{S. Miyabe$^{1,2}$ and P. Bucksbaum$^{1,3}$ }
\affiliation{$^1$Stanford PULSE Institute, SLAC National Accelerator Laboratory, 
Menlo Park, CA 94025\\
$^2$Department of Chemistry, Stanford University,
Stanford, CA 94305\\
$^3$Departments of Physics, Photon Science, and Applied Physics, Stanford University,
Stanford, CA 94305}
\begin{abstract}
Resonant Raman excitation by ultrafast vacuum ultraviolet
laser pulses  is a powerful means to study electron dynamics
in molecules, but experiments must contend with linear
background ionization: frequencies high enough to reach
resonant core-valence transitions will usually ionize all occupied
orbitals as well, and the ionization cross sections are usually
dominant.  Here we show that attosecond pulses can induce a 
process, transient impulsive stimulated Raman scattering, which
can overwhelm valence ionization. Calculations are performed
for atomic sodium, but the principal is valid for many molecular
systems. This approach opens the path for high fidelity
multidimensional spectroscopy with attosecond pulses.
\end{abstract}

\maketitle

High harmonic generation (HHG) sources and free electron lasers (FELs) can both produce pulses of short wavelength coherent radiation with duration $\tau$ on the order of a single femtosecond or less, and coherent bandwidth $\hbar / \Delta \tau$ of several $eV$ \cite{tanaka_proposal_2013,ding_generation_2009,hemsing_beam_2014,krausz_attosecond_2009,christov_high-harmonic_1997,agostini_physics_2004}. This is the energy spread associated with electronic structure.  Attosecond pulses therefore offer routes to study and potentially manipulate ultrafast electron dynamics of atoms and molecules on their intrinsic timescale.

Several experimental protocols have been proposed to study this new regime. Intense ultrashort pulses at extreme ultraviolet (XUV) or x-ray frequencies can excite a localized atomic core electron in a molecule to the valence levels.
This creates a coherent localized valence electronic wavepacket, which can be probed with additional pulses to map out the paths of ultrafast energy and coherence transport in molecular systems
\cite{:/content/aip/journal/jcp/136/17/10.1063/1.4706899,PhysRevLett.99.163001,li_population_2014,tanaka_coherent_2002}.
The potential applications of such coherent methods in molecules are vast \cite{Diddams19112004,2000:QCQ:544199,ShapiroBrumer2011}.

Such multidimensional spectroscopies must contend with the significant feature that core-valence transitions are embedded in the ionization continuum of the molecule. Therefore propagation of the excited electronic state generally occurs in the ion or dication rather than the neutral molecule. Raman scattering can transfer some population from the core states in the continuum back down to excited valence states. This is the principle that underlies resonant inelastic x-ray scattering (RIXS).  Unfortunately Raman cross sections are quite small so ionization is by far the dominant effect.
 
Coherent Raman methods such as Stimulated Rapid Adiabatic Passage (STIRAP) have led to efficient population transfer to excited states, but have severe limitations when the levels are broadened by coupling to the continuum\cite{Bergmann_AnnRev_2001}. More recently, multi-wavelength stimulated Raman methods and seeded Raman lasers have been proposed or demonstrated \cite{weninger_stimulated_2013,li_population_2014}.  

Here we carried out a detailed study of the transient impulsive electronic Raman transition efficiency initiated by high field attosecond XUV pulses and its competition with one-photon ionization.
We show that this process can create wave packets in neutral molecules, using atomic sodium as a test case.  Furthermore, we show that the electronic Raman redistribution has the remarkable property that for intense pulses of 1 fs or less it can overwhelm ionization to become the dominant process for electronic excitation in the molecule, and an efficient process for nonlinear spectroscopies.  Our calculations are done in neutral sodium and for pulses in the 30-40 eV region of the 2p-ns and 2p-nd autoionizing resonances; but the principle should be general and can be extended to 1s exitation by x-rays from next generation free electron lasers.
 
In the perturbative limit the scattering
differential cross section, involving initial state $m$ and final state $k$
can be obtained from Fermi's Golden rule and Kramers Heisenberg formula\cite{kramers_uber_1925,sakurai_advanced_1967}:
\begin{equation}
\label{eq:kh}
\frac{d\sigma}{d\Omega}= (N+1) \frac{\omega^3 \omega'}{c^4}
\left| \vec{\varepsilon} \cdot \alpha_{km} \cdot \vec{\varepsilon'} \right|^2
\end{equation}
where ${d\Omega}$ is the solid angle of the outgoing photon,
which is a  $\delta$-function along the photon propagation direction.
 Eq.~\ref{eq:kh} describes absorption of a photon
of frequency $\omega$' and emission of another at $\omega$.
${\vec{\varepsilon'}}$ and
${\vec{\varepsilon}}$ are the polarization vectors of the
incident and emitted photon, respectively,
and polarizability tensor ${\alpha_{km}}$ is
\begin{equation}
\label{eq:pol_tensor}
\left( \alpha_{km} \right)_{ij} =\frac{1}{\hbar}\sum_n
\left\{ -\frac{\mu^i_{nm}\mu^j_{kn}}{\omega-\omega_{nk}-i\Gamma} \right\},
\end{equation}
where $i$ and $j$ represent the polarization, ${\omega_{nk}}$ is
the energy  difference between the final and intermediate states, and
$\Gamma$ is the autoionization lifetime of the intermediate state.
Note that we are only interested in the polarized  scattering
event where ${i=j}$. The matrix elements are evaluated in the
dipole approximation. $N$ is the occupation number of the mode
responsible for stimulated emission in the Raman process given by
${N=\frac{I_{0}}{\omega^3 \alpha}}$. where ${\alpha}$ is the
fine structure constant and ${I_0}$ is the laser intensity per unit
bandwidth. Our pulse energy per unit area per bandwidth is
${F_0=\int dt I_0}$. To obtain the electronic Raman scattering probability $P$,
we multiply Eq.~\ref{eq:kh} by ${F_0}$ of the incident photon, and
integrate over the bandwidth of the pulse,
${P = \int dE \frac{d\sigma}{d\Omega} F_0}$.  

Eq.~\ref{eq:kh} is valid in the regime where $P \ll 1$. For higher intensities we need to consider the transient population dynamics of electronic Raman redistribution.  We utilized a simple time-dependent approach with a model Hamiltonian
applicable for
coherent laser
pulses with transform-limited bandwidths. We characterize the laser
field by its amplitude ${\mathcal{E}_0}$ and carrier frequency ${\omega}$, and
the electric field is $\vec{\mathcal{E}}(t)=\vec{\varepsilon}\;\mathcal{E}_{0}(t)\exp(-i\omega t)$.
 ${\vec{\varepsilon}}$ is the 
 polarization vector, and here we use linearly polarized light.
The state vector ${\left|\Psi(t)\right>}$ of an atom is expressed as a linear
combination of
electronic states that are involved in Raman transitions,
${\left|\Psi(t)\right>=\sum_{i}C_i(t)\exp(-iE_it/\hbar) \left| \phi_i \right>}$.
${C_i}$'s are time-dependent coefficients associated with electronic
state ${\left| \phi_i \right>}$ with energy ${E_i}$.

The expansion includes the initial ground
state, the intermediate excited states that are embedded in the continuum,
and the final valence excited states.
The final states we include in our expansion are 4s, 5s, 6s, 3d
and 4d states, and for each initial to final state transition we assume that
there is only one intermediate state. Thus, the intermediate states have
electronic configurations 2p$^5$3s($^1$P)ns or 2p$^5$3s($^1$P)nd.
We include the same states in the Kramers Heisenberg calculation.
For a three-level system, our model Hamiltonian reads
\begin{equation}
\label{eq:model_H}
\hat{H}(t)=
\left| \begin{array}{ccc}
E_{1}-i\gamma_{1}\:\:\:& d_{12}(t) & 0 \\
d_{21}(t) &  E_{2}-i\Gamma_{2} & d_{23}(t) \\
  0                       &  d_{32}(t) & E_{3}-i\gamma_{3}
  \end{array} \right|,
\end{equation}
where ${E_1}$, ${E_2}$ and ${E_3}$ are the energies of ground, intermediate
and final states, respectively. The off-diagonal coupling terms are given by
$d_{nm}(t)=\vec{\varepsilon}\cdot\vec{\mu}_{nm}\mathcal{E}_{0}(t)\exp(-i(\omega - \omega_{nm} t))$,
where $\vec{\mu}_{nm}$ and $\vec{\varepsilon}$ are defined as before.  
Here $\gamma$ describes the population loss due to one photon ionization, and $\Gamma$ includes both one photon ionization and autoionization of the intermediate state.
Thus, we have not included the resonantly enhanced two-photon ionization and the nonresonant Raman transitions through continua. 
We have computed the latter using the complex-Kohn scattering wavefunction in the perturbative limit and found that its probability is 0.3$\%$ of the resonant process.
We extend this three-level model to an n-level model by assuming
that the Raman transitions among final states are negligible. In a similar
manner, we ignore the two photon transitions among intermediate states.

\begin{figure}[h]
\includegraphics[width=0.68\columnwidth,clip=true]{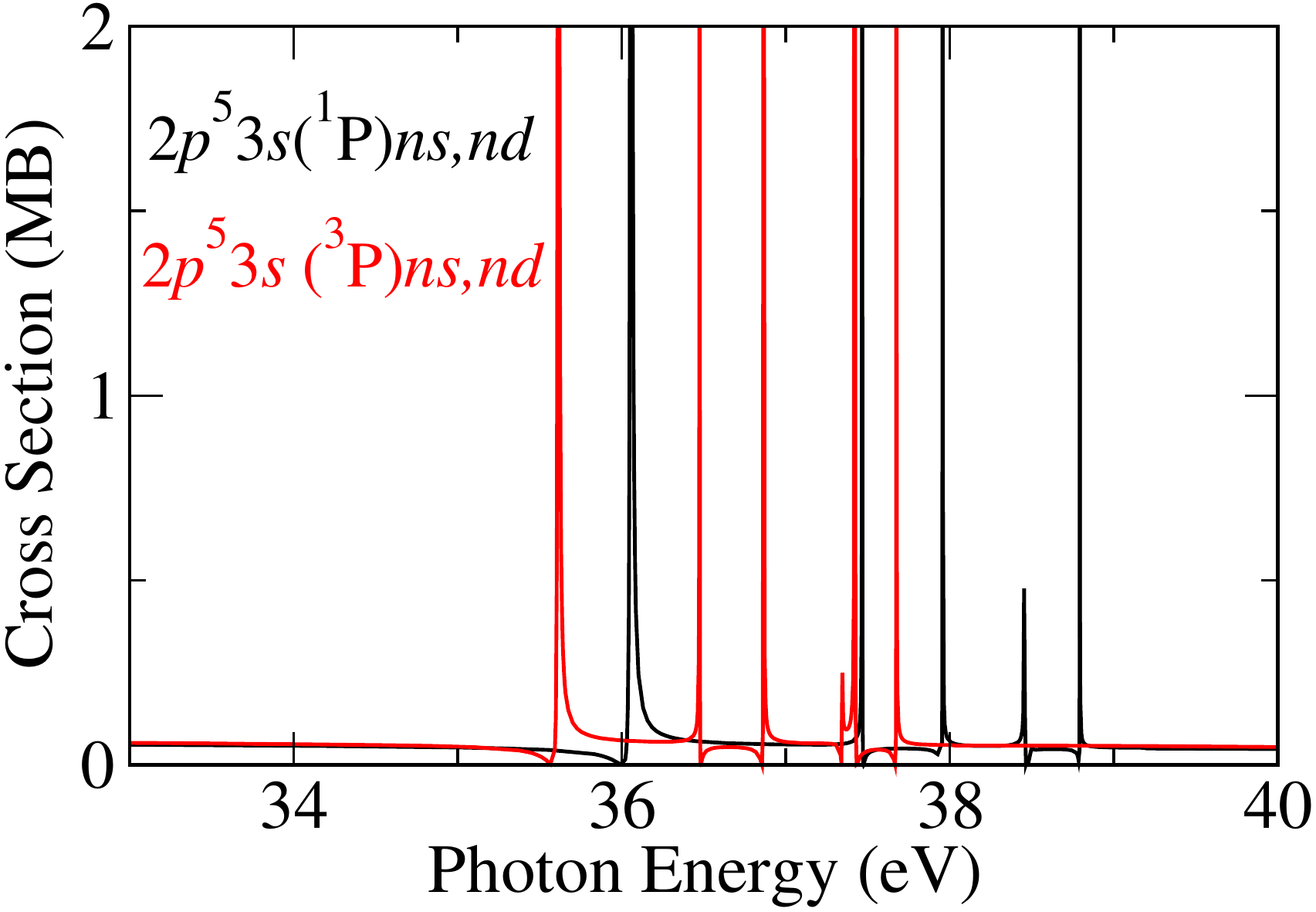}
\caption{(Color online) One photon single ionization cross section of
ground state Na in the resonance region. The black line shows the series
of resonance due to (${^1P}$)n$s$ or n$d$ states imbedded in the
continuum, and the red line is the resonances due to (${^3P}$)n$s$ or
n$d$ states } \label{fig:ionization}
\end{figure}

For both perturbative and non-perturbative approaches,
the transition matrix elements for bound-bound transitions are
computed using a complete active space self-consistent field (CASSCF)
method utilizing the Dunning triple zeta (TZP) basis set
\cite{doi:10.1080/00268970600899018} augmented
with four s-type, three p-type and two d-type diffuse functions. The exponents
are obtained from
\cite{0953-4075-22-14-007}. 
The relevant transition amplitudes from 3s to
ns final states were computed employing 8 electrons in 8 active
orbital (CASSCF(8/8)) approach. The active orbitals were 2p, 3s,
 3p, and ns. CASSCF(8/9) method, with active orbitals 2p, 3s,
 and nd were used for transitions ending in the nd states.
 Note that we used the Rydberg scaling law, which states that all 
core to Rydberg matrix elements scale as n$^{-3/2}$, to obtain the amplitudes involving the ns orbitals for n$>$4, and nd orbitals for n$>$3.
The relevant dipole transitions are shown in table.I.
The energies of
ground and excited states are obtained from NIST atomic spectra database
\cite{NIST_ASD}.
The single photoionization cross sections and Auger decay rates are
 computed using the scattering wavefunction obtained using the
 complex-Kohn variational method \cite{complex-kohn}. The continuum states were
expanded using the same Gaussian basis set described above, and
numerical continuum function up to ${l_{max}}$= 6 were included.
One-photon single ionization of 3s state in the 
resonance region is shown in Fig.~\ref{fig:ionization}. 
The autoionization rates are computed assuming that there is only one
intermediate state involved in a discrete transition from an initial to a
particular final state. The decay rates for the 5 intermediate states
are shown in table.II. 
Here we have once again used the Rydberg scaling law.
We use  their average value in computing the tensor
polarizability (Eq.~\ref{eq:pol_tensor}).

\begin{table}[h]
\begin{center}
\caption{Transition dipole moments calculated using the complete active
space self-consistent field method. See text for more detail.}
\begin{tabular}{l*{3}{c}r}
State 1  ~~~~ &  State 2~~~~ & dipole transition (au)  \\
\hline
3s              &  ($^1$P)4s &  0.247   \\
                 &  ($^1$P)5s  & 0.177  \\
                 & ($^1$P)6s  & 0.134  \\
                 & ($^1$P)3d  &  0.172 \\
                 & ($^1$P)4d &  0.112 \\
4s              &  ($^1$P)4s &  0.328\\
5s              &  ($^1$P)5s &  0.230\\
6s              &  ($^1$P)6s &  0.144\\
3d             &  ($^1$P)3d & 0.215\\
4d             &  ($^1$P)4d & 0.206\\
\end{tabular}
\end{center}
\label{table:dipole}
\end{table}
\begin{table}[h]
\begin{center}
\caption{Autoionization rates computed using the complex-Kohn variational method.}
\begin{tabular}{l*{2}{c}r}
Intermediate state  ~~~~ &  Autoionization rate (au)  \\
\hline
($^1$P)4s &  1.36e-5\\
($^1$P)5s &  6.96e-6\\
 ($^1$P)6s & 4.03e-6 \\
($^1$P)3d & 3.37e-4\\
($^1$P)4d & 1.41e-4\\
\end{tabular}
\end{center}
\label{table:auger}
\end{table}

\begin{figure}[h]
\includegraphics[width=0.7\columnwidth,clip=true]{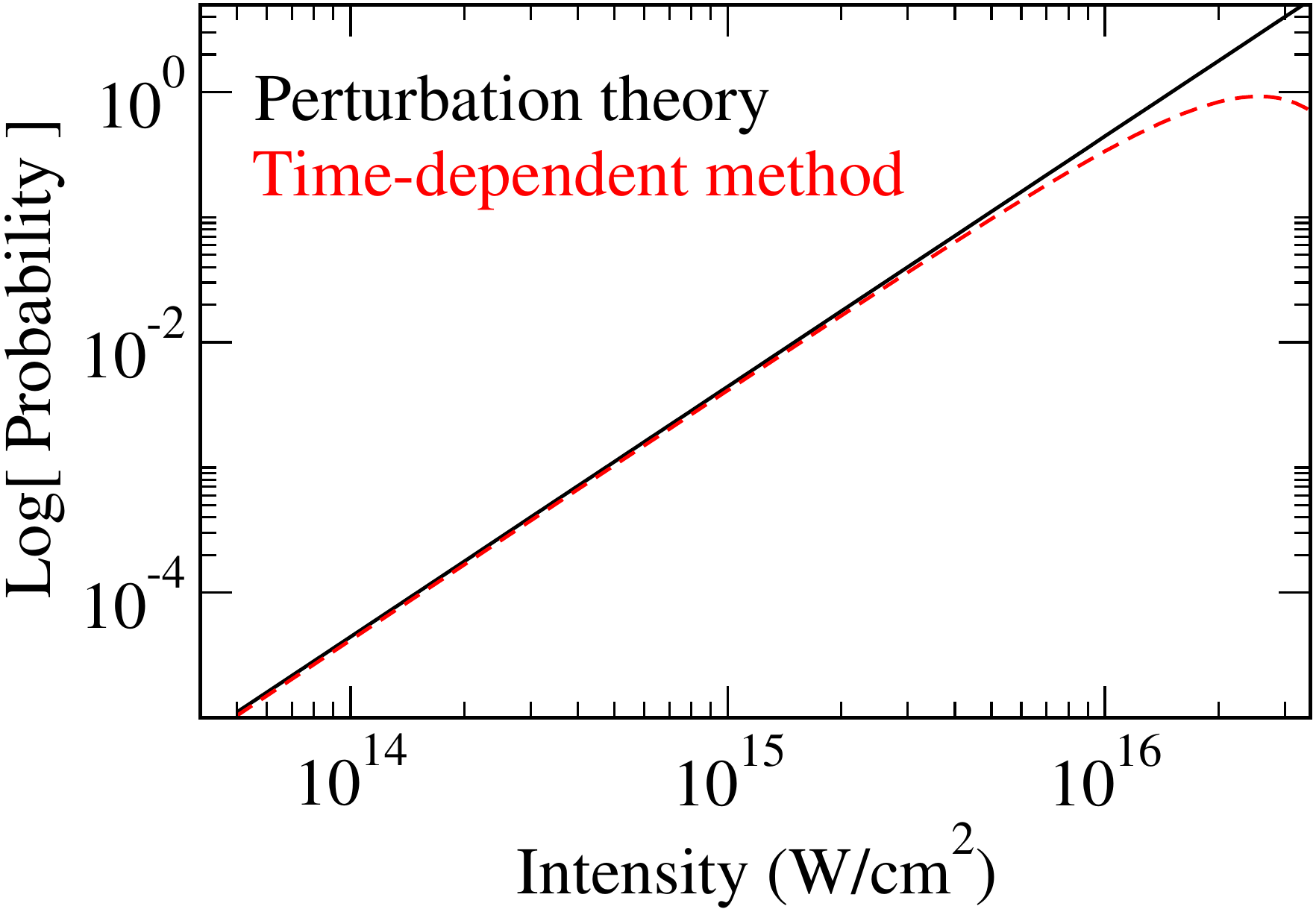}
\caption{(Color online) We plot the electronic Raman transition probability 
for Na atom as a function of maximum laser intensity (${I_0}$). The final state
population from a model time-dependent calculation is also shown. The pulse duration (FWHM) is
1fs and the central frequency is 35.0 eV.} \label{fig:raman_int}
\end{figure}

We plot the Raman probability as a function of maximum laser intensity in
Fig.~\ref{fig:raman_int}. 
Here the pulse duration (FWHM) is 1fs and the central frequency is 35.0 eV.
The transition probability increases quadratically in intensity and the time-dependent
calculation shows that the saturation limit is reached at ${I_0}$=2$\times
$10${^{16}}$W/cm$^2$. The perturbative result fails at high intensities
as expected. Both methods predict that $\sim$10$\%$
saturation limit is reached at ${I_0}$=5$\times$10${^{15}}$W/cm$^2$.
The dipole matrix elements coupling the initial and intermediate states are strong, 
and at the intensities considered here Rabi oscillations could occur and compete with the Raman
process. We have found that this effect becomes important for ${I_0}$$>$2$\times$10${^{16}}$W/cm$^2$.
This is evident from the figure, where we observe a decrease in the Raman probability above that intensity.

\begin{figure}[h]
\includegraphics[width=0.7\columnwidth,clip=true]{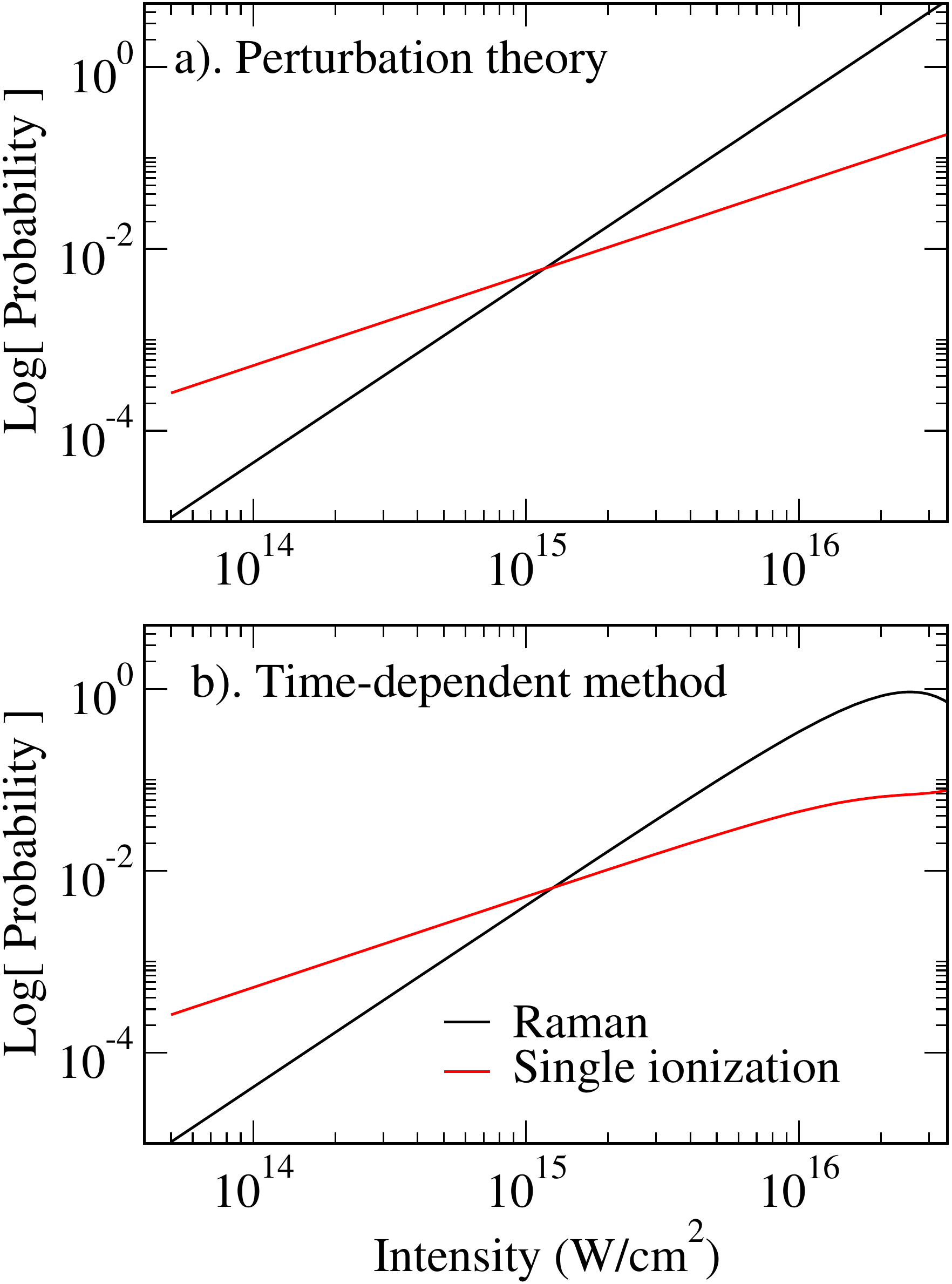}
\caption{We plot the Raman transition probability (black) and
one-photon single ionization probability (red) in Na atom as a function of maximum
laser intensity. The result obtained using
perturbation theory is shown. Pulse parameters are equal to 
 Fig.~\ref{fig:raman_int}} \label{fig:raman_v_ion}
\end{figure}

Fig.~\ref{fig:raman_v_ion}a compares the probability of the electronic Raman
transition to that of the
one-photon single ionization process obtained in the perturbative limit.
Note that here the probability for ionization
is obtained by taking the product of ${F_0}$ and the cross
section and integrating over the bandwidth of the pulse.
In Fig.~\ref{fig:raman_v_ion}b we compare
the two processes modeled by the time-dependent approach.
In both cases the parameter of the pulse is same
as the previous calculation.
Low intensity regime is
dominated by the direct ionization, while
for intensities greater than 10${^{15}}$ W/cm$^2$ the electronic
Raman transition starts to dominate.

\begin{figure}[h]
\includegraphics[width=0.8\columnwidth,clip=true]{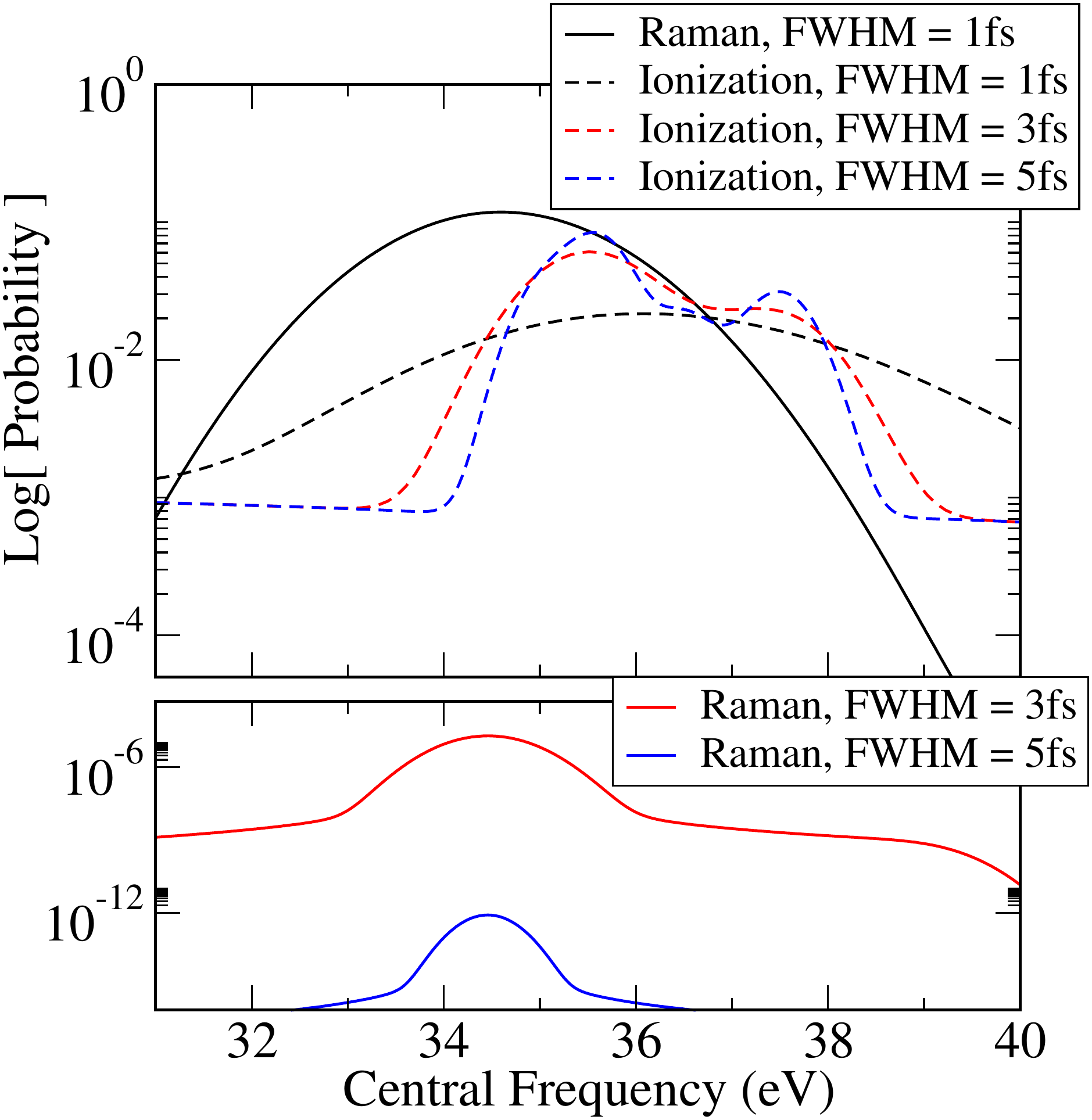}
\caption{We plot the Raman transition probability  and single ionization probability as a
function of the central frequency of the pulse for different pulse durations. $I_0$=5$\times$10$^{15}$ W/cm$^2$ for the 1 fs pulse, and the total pulse energy is kept constant for each pulse (see text). 
} \label{fig:raman_v_ion_cf}
\end{figure}

Fig.~\ref{fig:raman_v_ion_cf} shows the
 influence of pulse duration on the Raman process.
 Here, we focus on the perturbative calculation, and the maximum intensity is
 5$\times$ 10${^{15}}$ W/cm$^2$ for the 1 fs pulse.
 The total pulse energy (ie ${F_0}$ integrated over the bandwidth) is the same for each pulse.
 When the pulse duration is long ($\ge$ 3 fs) ionization overwhelms the Raman process.
 As the pulse duration is decreased, the Raman transition increases rapidly and
 for pulse duration $\sim$ 1fs it dominates the background ionization.

\begin{figure}[h]
\includegraphics[width=0.85\columnwidth,clip=true]{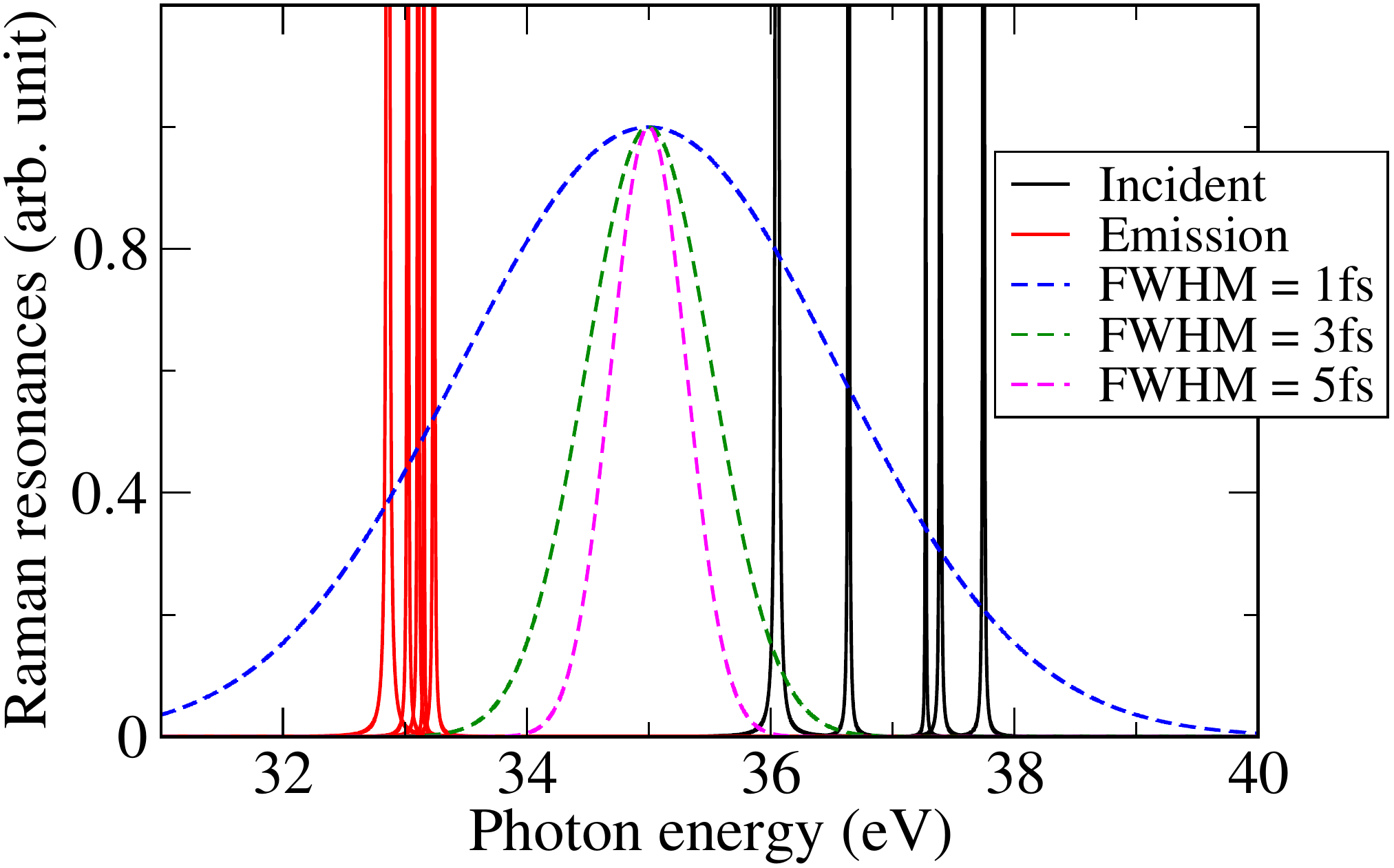}
\caption{The electronic Raman resonances for Na atom is plotted as a function
of both incident (black) and emitted (red) photon energies.
We also plot the energy bandwidth of the Gaussian shaped pulse envelope for
1fs, 3fs and 5 fs pulses. } \label{fig:raman_xsec_gaus}
\end{figure}

This rapid increase in the Raman probability becomes evident when its 
resonances (Eq.~\ref{eq:kh} without the $N+1$ factor) are plotted as a function of both incident and emitted photon energies
(Fig.~\ref{fig:raman_xsec_gaus}).
We also plot the energy bandwidth of our Gaussian-shaped pulse envelope,
centered at 35 eV, for 1 fs, 3 fs, and 5 fs pulses.
The Raman transitions become dominant when the energy bandwidth of the
pulse is large enough to efficiently cause both the
absorption (${\omega '}$) and stimulated emission (${\omega}$) events. When such
a condition is met, Raman redistribution overwhelms the ionization process
because Raman scattering increases quadratically with intensity, and because the 
Raman process can utilize the full bandwidth of the pulse. The Raman
resonances as a function of both incident and emitted
photon energies are plotted in Fig.~\ref{fig:raman_xsec_gaus}. 
3 fs and 5 fs can
 excite the atom from its ground state to an intermediate state, but they do not
 have enough bandwidth to cause an electronic Stokes-Raman transition. 1fs
 pulse, on the other hand, has enough bandwidth to
 cause the transition.

In conclusion, transient impulsive electronic Raman redistribution by ultrafast XUV laser pulse is a
powerful means to study electron dynamics in atoms and molecules. We have
shown that this technique can be used to create coherent
electronic valence wavepackets, thereby
imprinting the laser coherence directly on the electronic states which can
potentially be probed by other pulses. When the laser pulses are sufficiently short, Raman redistribution
can overwhelm photoionization. Furthermore, since core-level electrons
are involved in the transition, this technique provides highly selective
and localized view of electronic coherence. Current generation HHG sources should be able to access the intensity and pulse duration required for this process \cite{takahashi_attosecond_2013}.  The method should also work for more deeply bound 1s levels that can be excited by x-ray free electron lasers.  Transient impulsive electronic Raman redistribution creates a means to perform high fidelity studies of electron dynamics in molecules.

We gratefully acknowledge support from the Chemical Sciences Division of the SLAC National Accelerator
Laboratory by the US Department of Energy, Office of Basic Energy Sciences.
S. M. is grateful to Prof. C. W. McCurdy for helpful and inspiring discussions.

\bibliography{na_raman}

\end{document}


\bibliographystyle{apsrev}
\title{Transient Impulsive Electronic Raman Redistribution}
\author{S. Miyabe$^{1,2}$ and P. Bucksbaum$^{1,3}$ }
\affiliation{$^1$Stanford PULSE Institute, SLAC National Accelerator Laboratory,
Menlo Park, CA 94025\\
$^2$Department of Chemistry, Stanford University,
Stanford, CA 94305\\
$^3$Departments of Physics, Photon Science, and Applied Physics, Stanford University,
Stanford, CA 94305}

\maketitle

\section{Supplemental Material}

Atomic and molecular photoabsorption of electromagnetic pulses with frequencies high enough to reach resonant core-valence transitions will usually lead to ionization. However, here we showed that the transient impulsive giant electronic Raman redistribution can overwhelm ionization if intense pulses of 1 fs or less are used. In this section we discuss in detail the ionization processes that were considered in determining the likely occurrence of the Raman transitions.

The photoionization cross sections of the ns, nd, and the intermediate states were computed using the complex-Kohn variational method \cite{Rescigno2}. Here we give a brief summary. The final-state wave function for production of photoions in a specific cation state $\Gamma_0$ and with final angular momentum $l_0 m_0$ is written as
\begin{equation}
{\Psi^-_{\Gamma_0 l_0 m_0}=\sum_{\Gamma l m} A(\chi_\Gamma F^-_{\Gamma l m \Gamma_0 l_0 m_0}) +\sum_i d_i^{\Gamma_0 l_0 m_0} \Theta_i}
\end{equation}
where $\Gamma$ labels the  final ionic target states $\chi_\Gamma$ included,  {$F^-$} are channel functions that describe the photoionized electron, $A$ is the antisymmetrization operator, and the $\Theta_i$'s are $N$-electron correlation terms.
In the present application, we include 2p$^6$ and singlet (2p$^5$)ns or nd ionic states in the trial wavefunction. Note that for the intermediate state calculations we include only the singlet (2p$^5$)3s ionic state.

In the Kohn method, the channel functions are further expanded in the molecular frame as
\begin{equation}
\begin{split}
{F^{-}_{\Gamma l m \Gamma_0 l_0 m_0}}(\mathbf{r})  =
& \sum_i {c_i^{\Gamma l m \Gamma_0 l_0 m_0}}\varphi_i(\mathbf{r}) \\
&+
\sum_{lm}
\Big[
f_{lm}(k_\Gamma, {r})\delta_{ll_0}\delta_{mm_0} \delta_{\Gamma\Gamma_0} \\
&+
T_{ll_0mm_0}^{\Gamma\Gamma_0} h_{l m}^{-}(k_\Gamma, {r})
\Big]
Y_{lm}(\hat{\mathbf{r}})/{k_\Gamma^{\frac{1}{2}}r}\, ,
\end{split}
\label{eq:channelfcn}
\end{equation}
where the $\varphi_i(\mathbf{r})$ are a set of square-integrable (Cartesian Gaussian) functions,  {\it Y}$_{lm}$ is a normalized spherical harmonic, k$_{\Gamma}$ are channel momenta, and the $f_{lm}(k_\Gamma, {\mathbf r})$ and $h_{l m}^{-}(k_\Gamma, {\mathbf r})$ are numerical continuum functions that behave asymptotically as regular and incoming partial-wave Coulomb functions, respectively~\cite{Resc}. The coefficients T$_{ll_0mm_0}^{\Gamma\Gamma_0}$ are the T-matrix elements.

Photoionization cross sections in the molecular frame can be constructed from the matrix elements
\begin{equation}
{I^\mu_{\Gamma_0 l_0 m_0}=<\Psi^-_{\Gamma_0 l_0 m_0}}|r^\mu|\Psi_0>\, ,
\end{equation}
where $r^\mu$ is a component of the dipole operator, which we evaluate here in the length form,
\begin{equation}
r^\mu = \Bigg\{
\begin{array}{ll} z, &
\mu=0\\
\mp \left(x\pm i y\right)/\sqrt{2}, & \mu=\pm 1 \end{array}\,
\end{equation}
and $\Psi_0$ is the initial state wave function of the neutral $N$-electron target. In order to construct an amplitude that represents a photoelectron with momentum ${\bf k}_{\Gamma_0}$ ejected by absorption of a photon with  polarization direction $\hat{\epsilon}$, measured relative to the molecular body-frame, the matrix elements {$I^\mu_{\Gamma_0 l_0 m_0}$} must be combined in a partial wave series
\begin{equation}
\label{amplitude}
{I_{\hat{k} , \Gamma_0, \hat{\epsilon}}}=\sqrt{\frac{4\pi}{3}}\sum_{\mu l_0m_0} i^{-{l_0}}e^{i\delta_{l_0}} {I^\mu_{\Gamma_0 l_0 m_0}}
Y_{1\mu}(\hat{\epsilon})Y_{l_0m_0}{(\hat{k})}\, ,  
\end{equation}
where $\delta_{l_0}$ is a Coulomb phase shift. The differential cross section with respect to the angle of photoejection and photon polarization in the fixed body-frame of the molecule is then given by
\begin{equation}
\frac{d^2\sigma}{d\Omega_{\hat{k},{\Gamma_0}} d\Omega_{\hat{\epsilon}}}=
\frac{8\pi \omega}{3c}{|I_{\hat{k},{\Gamma_0},\hat{\epsilon}}|^2}\, ,
\end{equation}
where $\omega$ is the photon energy and $c$ is the speed of light.

The photoionization cross section of the ground state was computed using the Hartree-Fock orbitals of the neutral atom. To determine the cross section of ns and nd states we start withe a reference configuration with a vacancy in the 3s orbital, but a single electron occupancy in the ns or nd orbital. We then perform an all-singles configuration-interaction (CIS) calculation, keeping the 3s occupancy zero. The natural orbitals from that calculation, obtained by diagonalizing the one-particle density matrix, are then used in the photoionization calculations. A similar procedure was used to compute the ionization amplitudes of the intermediate states. We start with a reference configuration of (2p$^5$)3s ns or nd state, and we then carried out a CIS calculation to obtain orbitals corresponding to the relevant excited state configuration. In our time-dependent methods we use the cross section averaged over the bandwidth of the pulse.

\begin{figure}[h]
\includegraphics[width=0.6\columnwidth,clip=true]{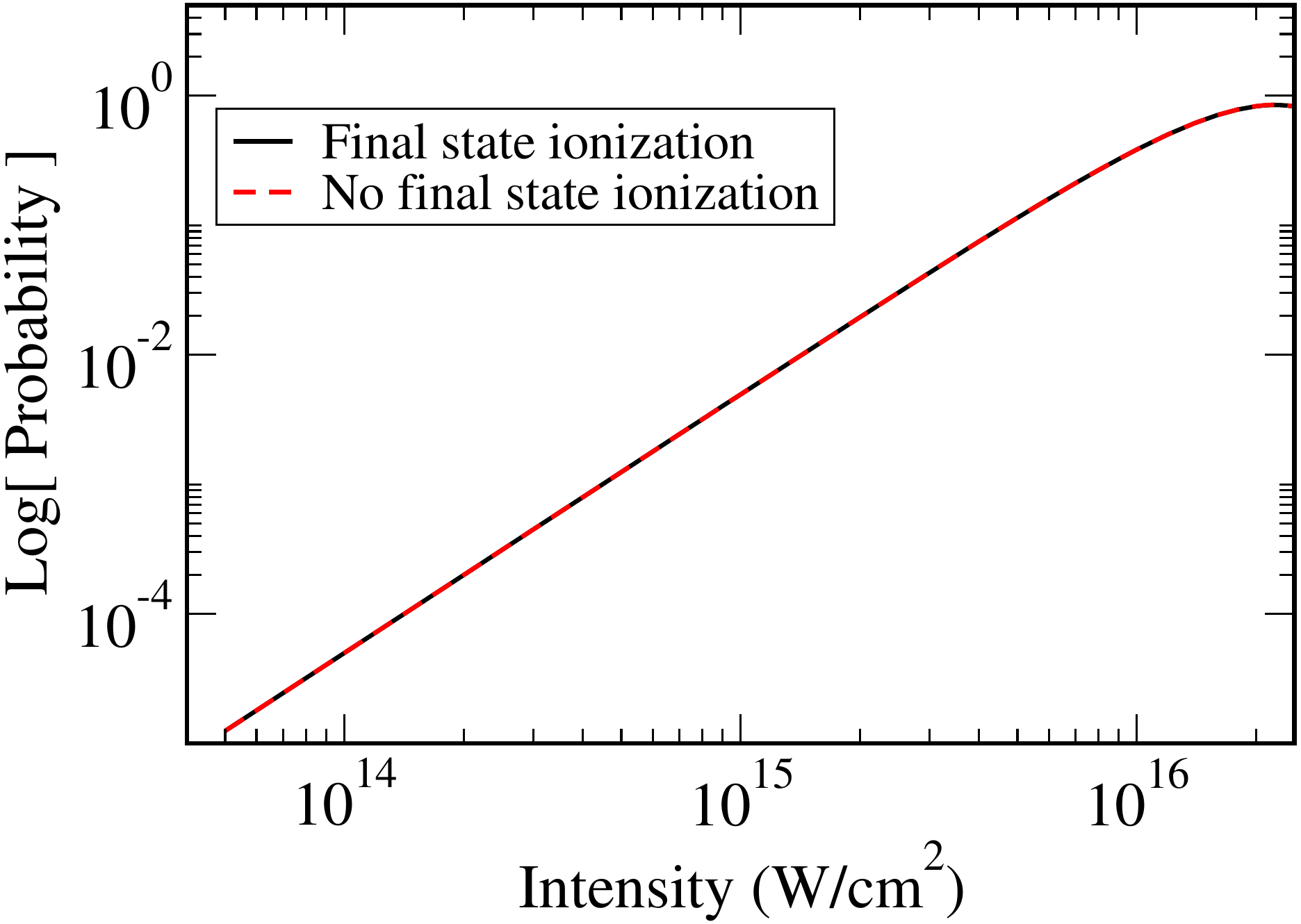}
\caption{(Color online) We plot the Raman transition probability in Na atom as a function of maximum intensity. (Black) The calculation was carried out with the inclusion of population loss due to final state photoionization. (Red) For comparison we also carried out our simulation without considering the final state ionization.} \label{fig:final_state}
\end{figure}

The process that competes strongly with the Raman transition is the single-photoionization of the ground state, and this has been discussed in detail in the main text. We also considered a population loss due to final state photoionization. We included this effect by adding the ionization rate as an imaginary component of the diagonal matrix matrix elements (See Eq. 3 of main text). In Fig.~\ref{fig:final_state} we plot the Raman transition probability as a function of intensity. Here the pulse duration is 1fs and the central frequency is 35.0 eV. We see that the two calculations, with and without the final state ionization contribution, are graphically indistinguishable.

\begin{figure}[h]
\includegraphics[width=0.6\columnwidth,clip=true]{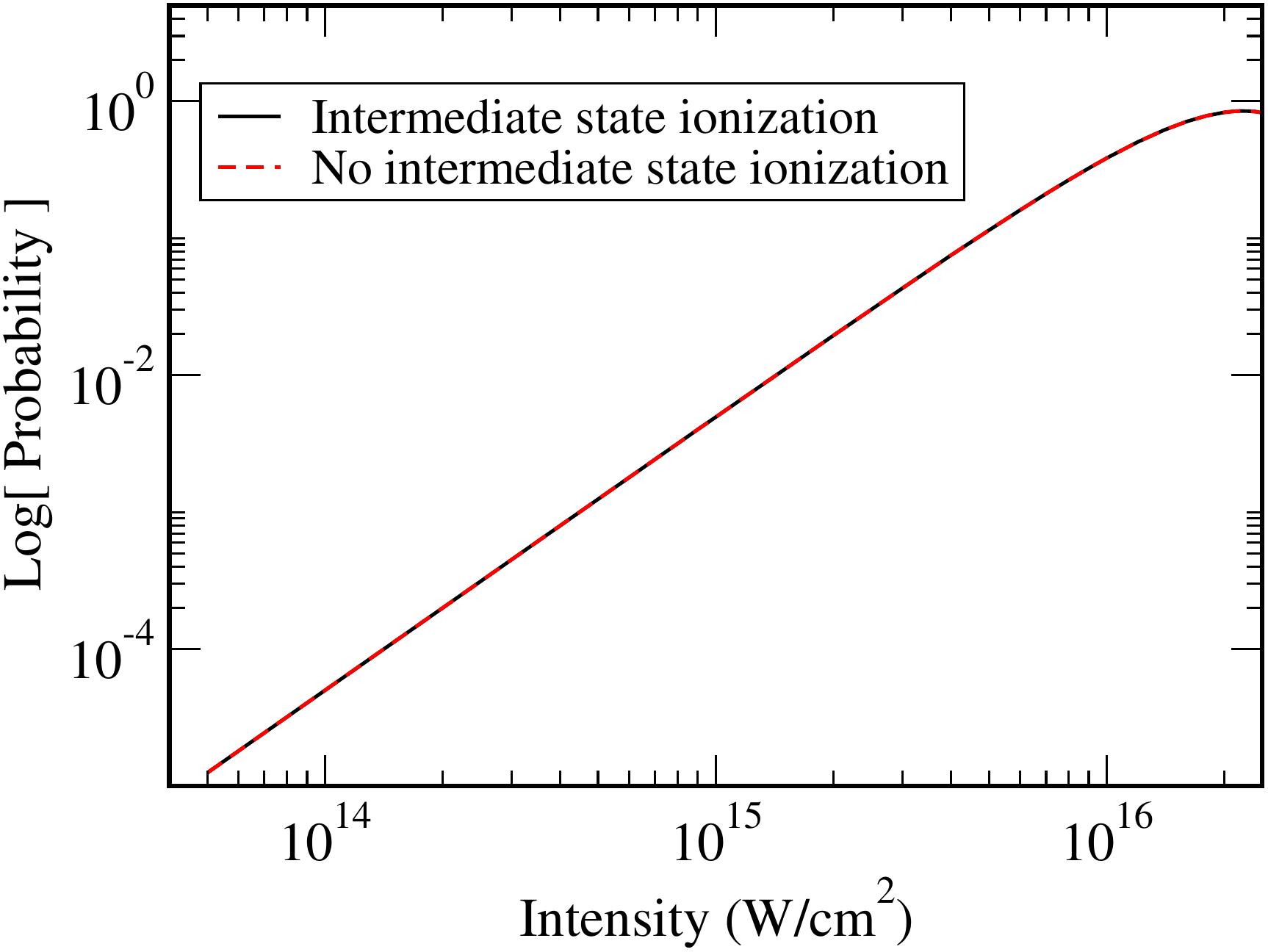}
\caption{(Color online) We plot the Raman transition probability in Na atom as a function of maximum intensity. (Black) The calculation was carried out with the inclusion of population loss due to intermediate state photoionization. (Red) For comparison we also carried out our simulation without considering the process.} \label{fig:inter_state}
\end{figure}

One-photon ionization of intermediate states is another process that competes with the Raman yield. At the pulse intensities that we consider photoionization rates are much larger than the autoionization rates, and this might have an important consequence on the Raman probability. Thus, we have added these rates on the diagonal matrix elements corresponding to the intermediate states. We once again carried out the model time-dependent simulation with and without the inclusion of this contribution and the result is shown in Fig.~\ref{fig:inter_state}. We see that the two calculations are identical within the graphical accuracy.

\begin{figure}[h]
\includegraphics[width=0.6\columnwidth,clip=true]{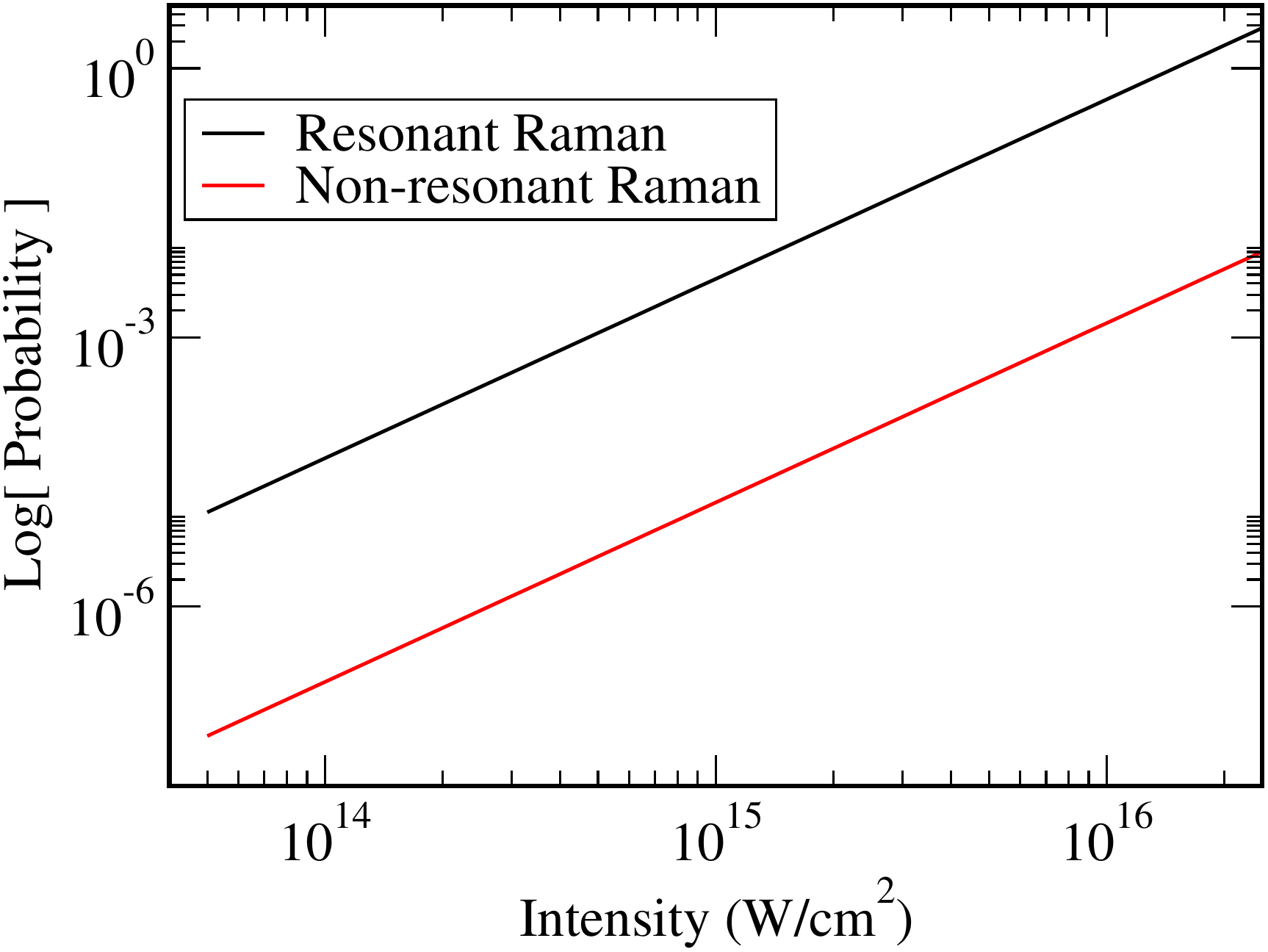}
\caption{(Color online) We plot the resonant (Black) and non-resonant (Red) Raman transition probability in Na atom as a function of maximum intensity.} \label{fig:res_nonres}
\end{figure}

Our main focus here was on Raman transitions from the ground state to valence excited states through bound autoionizing intermediate states. The Raman transitions can also occur through continuum states in the vicinity of such quasi-bound intermediate states. We therefore computed the non-resonant Raman transition probability. We plot the transition probability for both resonant and non-resonant Raman transitions in Fig.~\ref{fig:res_nonres}. We found that the latter effect is roughly 0.3 \% of the resonant process, having negligible effect on the overall Raman probability

\begin{figure}[h]
\includegraphics[width=0.6\columnwidth,clip=true]{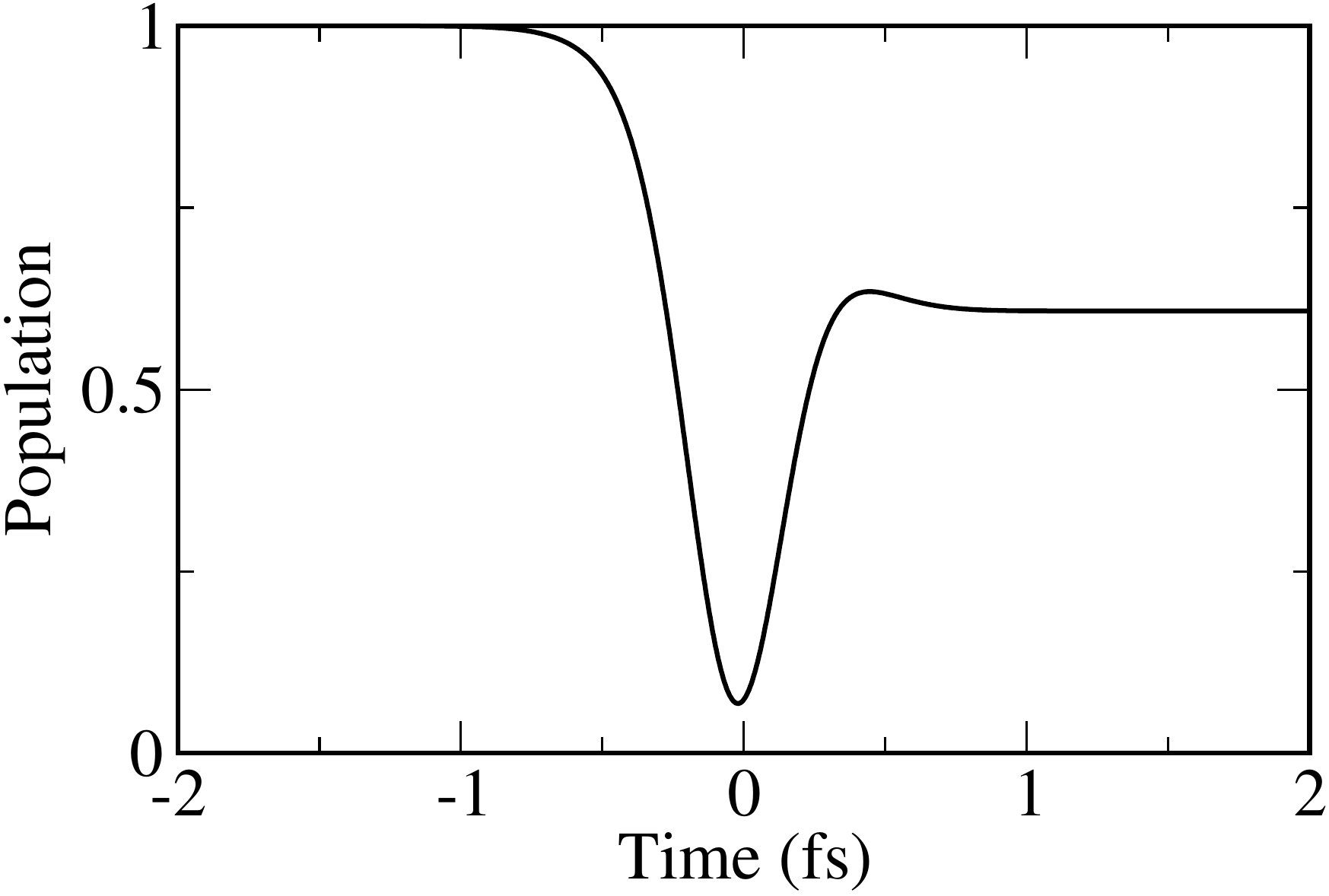}
\caption{(Color online) We plot the ground state population from our model time-dependent simulation. Here the intensity is 5$\times$10$^{16}$W/cm$^2$. The pulse duration is 1fs and the central frequency is 35.0 eV.} \label{fig:rabi}
\end{figure}

Finally, there is a strong dipole coupling between the initial and intermediate autoionizing states. Thus, Rabi oscillations should take place at the intensities considered in this paper. We see that such effect becomes important at intensities greater than 2$\times$10$^{16}$W/cm$^2$. In Fig.~\ref{fig:rabi} we plot the ground state population in our time-dependent simulation. Here the intensity is 5$\times$10$^{16}$W/cm$^2$. The pulse duration is 1fs and the central frequency is 35.0 eV. We observe a depletion of the ground state population starting around -0.2 fs, but the population comes back to the ground state starting at 0 fs.

